\documentclass[runningheads]{llncs}
\usepackage[T1]{fontenc}
\usepackage{graphicx}
\usepackage{amsmath}
\usepackage{paralist, blindtext}
\usepackage{amssymb}
\newcommand {\pf}{\sc Proof.\ }
\newcommand {\epf}{$\dashv$}
\newcommand {\ra}{\rightarrow}

\newcommand{\badstart}[0]{\ \\[-.2in]}

\newcommand \infer[2]{\begin{array}{c}#2\\ \hline#1\end{array}}

\newcommand{\wbg}[1]{{[{#1}]_{\mathbf o}}}
\newcommand{\wbm}[1]{[{#1}]_{\mathbf p}}

\newcommand{\wnbg}[1]{[\![ {#1}]\!]_{\mathbf o}}
\newcommand{\wnbm}[1]{[\![ {#1}]\!]_{\mathbf p}}

\newcommand{\wnb}[1]{[\![ {#1}]\!]}

\newtheorem{mylma}{Lemma}
\newtheorem{myex}{Example}
\newtheorem{mycor}{Corollary}

\usepackage{latexsym}
\DeclareGraphicsExtensions{.jpg,.pdf,.mps,.png}
\usepackage[colorlinks=true, allcolors=blue,unicode]{hyperref}
\usepackage{color}

\begin{document}
\title{A Modal Logic for Possibilistic Reasoning with Fuzzy Formal Contexts}
%
%
\author{Prosenjit Howlader
\and Churn-Jung Liau\thanks{The corresponding author}
}
\authorrunning{P. Howlader and C.J. Liau}
%
\institute{Institute of Information Science, Academia Sinica,  Taipei 115, Taiwan 
\\
\email{\{prosen,liaucj\}@iis.sinica.edu.tw}}
\maketitle              
\begin{abstract}
We introduce a \emph{two-sort weighted modal logic} for possibilistic reasoning with fuzzy formal contexts. The syntax of the logic includes two types of weighted modal operators corresponding to classical \emph{necessity} ($\Box$) and \emph{sufficiency} ($\boxminus$) modalities and its formulas are interpreted in \emph{fuzzy formal contexts} based on possibility theory. We present its axiomatization that is \emph{sound} with respect to the class of all fuzzy context models. In addition, both the necessity and sufficiency fragments of the logic are also individually \emph{complete} with respect to the class of all fuzzy context models.  We  highlight the expressive power of the logic with some illustrative examples. As a formal context is the basic construct of formal concept analysis (FCA), we generalize three main notions in FCA, i.e., \emph{formal concepts}, \emph{object oriented concepts}, and \emph{property oriented concepts}, to their corresponding $c$-cut concepts in fuzzy formal contexts.  Then, we show that our logical language can  represent all three of these generalized notions. Finally, we demonstrate the possibility of extending our logic to reasoning with multi-relational fuzzy contexts, in which the Boolean combinations of different fuzzy relations are allowed.

\end{abstract}
\section{Introduction}
Formal concept analysis (FCA) is a mathematical theory for studying structures of concepts~\cite{FCA}. In FCA, a {\em formal context} (or simply context), aka {\em polarity\/}, is a triple $(G,M,I)$, where $G$ and $M$ are sets of objects and attributes, respectively, and $I \subseteq G \times M$ is a binary incidence relation connecting them. Given a context, a {\em formal concept\/} is a pair of subsets of objects and attributes $(A,B)$ such that $A$ is the set of objects possessing all attributes in $B$ and $B$ is the set of attributes shared by all objects in $A$. For the formal concept, $A$ and $B$ are called its extent and intent, respectively. Furthermore, inspired by rough set theory~\cite{pawlak2012rough}, notions of rough concepts including object and property oriented concepts, have been also defined~\cite{duntsch2002modal,yao2004comparative,yao2004concept}.

The theoretical framework of FCA has evolved into an indispensable tool for knowledge discovery  from complex datasets which are usually presented as formal contexts . Traditionally, most data mining methods represent the discovered knowledge as simple if-then rules. However, this is not expressive enough for the representation of more complicated knowledge. Moreover, when applying to decision-making, simple rules can facilitate only shallow reasoning by matching their antecedents with the decision situation and taking consequents of matched rules. By employing a logical approach, we can address the expressivity of knowledge representation formalisms for data mining from formal contexts. 

For the knowledge representation and reasoning with formal contexts, several logical formalisms have received much attention~\cite{dunn,CWFSPAPMTAWN,Conradie2017167,HCgame,CH,HOWLADER2023115,ijcrs23}. In a recent work, it is shown that  two-sorted Boolean modal logic can serve as a reliable base for the representation of both formal and rough concepts~\cite{bmlpl}. The logic is a synergy of two-sorted modal logic and Boolean modal logic. 

On one hand, two-sorted modal logic is an instance of many-sorted polyadic modal logic, which was studied in \cite{mspml} and then extended to a many-sorted hybrid logic in \cite{LEUSTEAN}. Because a formal context has two domains and an incidence relation, it is an appropriate semantic model for two-sorted modal logic. On the other hand, Boolean modal logic combines two dual branches of modal logic: one with the box modality $(\square)$ representing necessity, and the other with the window modality $(\boxminus)$ representing sufficiency~ \cite{Gargov1987}. These two kinds of modalities correspond to basic operators in rough approximations and FCA, and thus facilitate the representation of different notions of concepts in formal contexts.

While FCA has been an effective tool for data analysis, because of the pervasive uncertainty and vagueness of information in complex datasets, the binary relation in formal contexts may be not two-valued in many applications. In particular, to deal with vague information in formal contexts, various fuzzy FCA methods have been proposed~\cite{Antoni2018,Belohlavek04,Belohlavek11,BelohlavekV05,BritoBEBC18}, in which a formal context is extended to a fuzzy context by replacing the crisp incidence relation with a fuzzy one.  The derivation operators map fuzzy sets on $G$ to fuzzy sets on $M$ and vice versa, forming a Galois connection. Consequently, fuzzy concepts are pairs of fuzzy sets mutually determined by these operators. Typically, membership functions of these fuzzy sets measure the degree to which an attribute (object) connects with all objects (attributes). However, unlike the existing approaches, we further define the derivation operators using (strict) $c$-cuts for $c \in [0,1]$ on these fuzzy sets, mapping a set of objects to the attributes whose degree of connecting with all these objects meets or exceeds $c$, and vice versa.

As logics for fuzzy formal contexts are under-explored yet, there exist only a few works on this topic. Among them,  two approaches to connecting logic with fuzzy formal contexts have been proposed in \cite{CodaraEGV18}. One  is to follow an idea in \cite{CodaraV15} by interpreting fuzzy logic formulas as formal concepts, and the other is to consider semantic evaluations and formulas of a fuzzy logic as objects and attributes of a fuzzy context, respectively, and use the fuzzy relation between them to denote the satisfaction degree of a formula in an evaluation.
	
In this paper, we present yet another approach based on the weighted extension of the above-mentioned two-sorted Boolean modal logic. By considering the fuzzy incidence relation as a class of possibility distributions, we use weighted modalities to represent uncertainty measures corresponding to different operators in FCA as explicated in \cite{DuboisSP07}. In addition, as these weighted modalities correspond to our cut-based derivation operators,  we define $c$-cut concepts as the generalization of formal and rough concepts in fuzzy contexts and show that our logic can represent and reason with these generalized  concepts.

The remainder of the paper is organized as follows. The next section contains background knowledge on FCA,  fuzzy sets, and possibility theory.  In Section~\ref{sec3}, we introduce basic operators for constructing generalized concepts in fuzzy formal contexts. In addition, we define the cut-based notions of formal and rough concepts. In  Section~\ref{sec4}, we present the two-sorted weighted modal logic and show its application to the representation of different $c$-cut concepts. While the completeness of the proposed logic remains an open issue, we prove that its necessity and sufficiency fragments are individually complete in  Section~\ref{sec5}. In Section~\ref{sec6}, we discuss the possible extension of the logic to reasoning with multi-relational fuzzy contexts.  We conclude the paper with some directions of future work  in Section~\ref{sec7}.

\section{Preliminaries}\label{sec2}
In this section, we introduce some preliminary notions about FCA and possibility theory.
\subsection{Formal concept analysis}
Let us recall the definition of a context $(G, M, I)$ introduced in the previous section.
A given context induces two derivation operators $+: (\mathcal{P}(G), \subseteq)\rightarrow (\mathcal{P}(M), \supseteq) $ and $-: (\mathcal{P}(M), \supseteq)\rightarrow (\mathcal{P}(G), \subseteq)$, where for all $A\in \mathcal{P}(G)$ and $B\in \mathcal{P}(M)$:

$A ^{+}= \{m\in  M\mid\forall g(g\in A\Rightarrow gIm)\},~~~B ^{-}= \{g\in  G\mid\forall m(m\in B\Rightarrow  gIm)\}.$

It can be shown that the pair of maps forms a Galois connection. Moreover, every Galois connection is associated with some formal context, and this correspondence is bijective. For details, we refer the reader to \cite{textfca}.

 A {\it (formal) concept} is a pair of sets $(A, B)$ such that $A^{+}=B$ and $A=B^{-}$, where $A$ and $B$ are called its {\it extent } and {\it intent}, respectively. We denote the set of all concepts by $\mathbf{B}(\mathbb{K})$. It forms  a complete lattice and is denoted by $\underline{\mathbf{B}}(\mathbb{K})$.

D\"{u}ntsch and Gediga \cite{duntsch2002modal} defined sufficiency, dual sufficiency, possibility and necessity operators based on a context. In particular, for a context  $\mathbb{K}:=(G,M,I)$,  $A\subseteq G$, and $B\subseteq M$, the pairs of possibility and necessity operators are defined as follows:

$A^{\Diamond}:=\{m\in M\mid\exists g(gIm\wedge g\in A)\},~~~ A^{\Box}:=\{m\in M\mid\forall g(gIm\Rightarrow  g\in A)\}$.

$B^{\Diamond}:=\{g\in G\mid\exists m(gIm\wedge m\in B)\},~~~~B^{\Box}:=\{g\in G\mid\forall m(gIm\Rightarrow m\in B)\}$.

These operators correspond to approximation operators in rough set theory (RST) \cite{pawlak1982rough}. Based on them, D\"{u}ntsch and Gediga \cite{duntsch2002modal} and Yao \cite{yao2004concept}  introduced property oriented concepts and object oriented concepts, respectively.
A pair $(A,B)$ is a  {\it property oriented concept}  of  $\mathbb{K}$ iff $ A^{\Diamond}=B$ and $B^{\Box}=A$, and it  is an {\it object oriented concept}  of $\mathbb{K}$ iff $ A^{\Box}=B$ and $B^{\Diamond}=A$. These two kinds of concepts are also called rough concepts

In addition to formal concept lattices mentioned above, the set $\mathbf{O}(\mathbb{K})$ of all object oriented concepts and the set $\mathbf{P}(\mathbb{K})$ of all property oriented concepts also form complete lattices,  which are called {\it object oriented concept lattices} and {\it property oriented concept lattices}, respectively. The relationship among the three kinds of concept lattices are investigated in \cite{yao2004comparative}.

Regarding the motivation of the paper, we use the following example to show how a formal context can represent a dataset in practical applications.
	\begin{myex}\label{ex1}{\rm
		Let us consider a formal context $(G,M,I)$, where $G$ is the set of readers (users) of an on-line document repository $M$ containing books, articles, and so on. In such a scenario, for  $g\in G$ and $m\in M$, $gIm$ means that the reader $g$ has downloaded the document $m$. Assume that $A$ is a particular group of users and $B$ is a special class of documents. Then, the above-mentioned operators of FCA may contain the following information:
		\begin{itemize}
			\item $g\in B^\Box$:  the reader $g$ only downloaded documents in $B$
			\item $m\in A^\Box$: only readers in $A$ downloaded the document $m$
			\item $g\in B^-$: the reader $g$ downloaded all documents  in $B$
			\item $m\in A^+$: all readers in $A$  downloaded the document $m$
		\end{itemize}}
	\end{myex}

\subsection{Fuzzy set, possibility theory, and possibilistic reasoning}
Fuzzy set theory is invented by Zadeh~\cite{zad0} to model vagueness in set membership. A fuzzy set is a set without a precise boundary between its elements and non-elements. Given a universe $W$, a fuzzy set $\tilde{A}$ over $W$ is defined as a membership function $\tilde{A}:W\ra[0,1]$. We denote the class of all fuzzy subsets of a universe $W$ by $\tilde{\mathcal P}(W)$. A fuzzy binary relation between two domains $U$ and $V$ is simply a fuzzy subset of its Cartesian product $U\times V$. Let $\tilde{A}$ be a fuzzy set over $W$ and $c\in [0,1]$. Then, the $c$-cut and strict $c$-cut of $\tilde{A}$ are defined as $\tilde{A}_c=\{w\in W\mid\tilde{A}(w)\geq c\}$ and $\tilde{A}_{>c}=\{w\in W\mid\tilde{A}(w)> c\}$, respectively. 

When the universe $W$ is a set of possible worlds (or states), we can interpret a fuzzy set as a {\em possibility distribution \/}. This leads to the development of possibility theory~\cite{zad}. 
	
\begin{definition}{\rm\cite{DuboisSP07}
Let $W$ be a set. Then, we define
		\begin{itemize}
			\item   a possibility distribution  over $W$ as a function $\pi:W\ra[0,1]$,
			\item the possibility measure  $\Pi:\mathcal{P}(W)\rightarrow[0,1]$ by $\Pi(X):= \sup_{x\in X}\pi(x)$, 
			\item the necessity measure $N:\mathcal{P}(W)\rightarrow[0,1]$ by $N(X):= \inf_{x\not\in X}(1-\pi(x))$,
			\item the guaranteed possibility measure  $\vartriangle:\mathcal{P}(W)\rightarrow[0,1]$ by $\vartriangle\!\!(X) :=\inf_{x\in X}\pi(x)$, 
			\item the potential certainty measure  $\triangledown:\mathcal{P}(W)\rightarrow[0,1]$ by $\triangledown(X):=\sup_{x\not\in X}(1-\pi(x))$.
		\end{itemize}}
	\end{definition}
By mathematical convention, $\sup\emptyset=0$ and $\inf\emptyset=1$. Hence, by the definition, we have $\Pi(\emptyset)=\triangledown(W)=0$	and $N(W)=\vartriangle\!\!(\emptyset)=1$.

Possibilistic logic has been proposed for reasoning about possibility distributions~\cite{dp88,dp94}. By the analogy between necessity (resp. possibility) measure and the modal operator $\Box$ (resp. $\Diamond$) indicated in \cite{dp88}, more expressive modal formulations of possibilistic reasoning have been also investigated in \cite{dc,qml1,qml2,qml4}. The necessity fragment of the logic to be introduced below will be the two-sorted version of the QML proposed in \cite{qml1}.
	
\section{Fuzzy Formal Context and Cut-based Concepts}\label{sec3}
	Let us recall the scenario in Example \ref{ex1}. Sometimes, the data may contain vague or more fine-grained information. For example, it may be known that how  strongly a reader is interested in a document.  To model such situations, we introduce the fuzzy formal context, which is a special instance of formal {\bf L}-contexts defined in \cite{Belohlavek04}, where {\bf L} is any residuated lattice.
    \begin{definition}
        {\rm A {\em fuzzy formal context} is defined as $(G, M, \tilde{I})$, where $G$ and $M$ are the same as above, but $\tilde{I}:G\times M\ra[0,1]$ is now a fuzzy binary relation.}
    \end{definition} 
    Intuitively, $\tilde{I}(g,m)$ specifies the strength or intensity of the connection between $g$ and $m$.
	
Given a fuzzy formal context $(G,M,\tilde{I})$, we can define possibility distributions $\pi_g:M\ra[0,1]$ for each $g\in G$ and $\pi_m:G\ra[0,1]$ for each $m\in M$
	 in the following way:
	\[\pi_g(m)=\tilde{I}(g,m),~m\in M\]
	\[\pi_m(g)=\tilde{I}(g,m),~g\in G.\] 
The definitions of $(\cdot)^+$, $(\cdot)^-$, and  approximation operators in fuzzy formal context are similar to those for formal context except that the incidence relation $I$ is replaced by $\tilde{I}$ and logical connectives and quantifiers are replaced by those for fuzzy logic. Formally, we can  use the residuum of {\L}ukasiewicz t-norm~\cite{haj} to define these operators. As the {\L}ukasiewicz t-norm $*:[0,1]^2\ra[0,1]$ is simply $a*b=\max(a+b-1,0)$, its residuum $\Rightarrow:[0,1]^2\ra[0,1]$ is defined as $a\Rightarrow b=\sup\{c\mid a*c\leq b\}$. By the definition, we have $a\Rightarrow b=1$ if $a\leq b$ and $a\Rightarrow b=1-a+b$ if $a> b$. Then, we can define corresponding operators in fuzzy formal contexts as follows.

	\begin{definition}\label{fuzzyset}{\rm	Let $( G, M, \tilde{I})$ be a fuzzy  formal context and let $A\in \mathcal{P}(G)$ and $B\in \mathcal{P}(M)$. 
		Then, we define membership functions of fuzzy sets $A ^{+}, A^\Box, A^\Diamond, B^-,  B^\Box$, and $B^\Diamond$ by the following equations.
		\begin{equation}A ^{+}(m):=\inf_{g\in G}(\chi_A(g)\Rightarrow\tilde{I}(g,m))\end{equation}
		\begin{equation}B ^{-}(g):=\inf_{m\in M}(\chi_B(m)\Rightarrow\tilde{I}(g,m))\end{equation}
		\begin{equation} A^\Box(m):=\inf_{g\in G}(\tilde{I}(g,m)\Rightarrow\chi_A(g))\end{equation}
		\begin{equation} B^\Box(g):=\inf_{m\in M}(\tilde{I}(g,m)\Rightarrow\chi_B(m))\end{equation}
		\begin{equation} A^\Diamond(m):=\sup_{g\in G}\min(\tilde{I}(g,m),\chi_A(g))\end{equation}
		\begin{equation} B^\Diamond(g):=\sup_{m\in M}\min(\tilde{I}(g,m),\chi_B(m))\end{equation}
		where $\chi_A$ and $\chi_B$ are the characteristic functions of $A$ and $B$, respectively.}
	\end{definition}
Intuitively, $A^+(m)$ denotes the degree of $m$ being an attribute of all objects in $A$, whereas $B^-(g)$ means the degree of $g$ being an object with all properties in $B$. On the other hand, $A^\Box(m)$ is the degree that all objects with attribute $m$ belong to $A$ and $B^\Box(g)$ denotes the degree that all attributes of $g$ are in $B$. In terms of rough set theory, $A^\Box(m)$ (resp. $B^\Box(g)$) denotes the membership degree of $m$ (resp. $g$) in the lower approximation of $A$  (resp. $B$). Analogously, $A^\Diamond$ and $B^\Diamond$ correspond to upper approximations of $A$ and $B$ in fuzzy contexts, respectively.

Note that the operators above are defined with respect to a  fuzzy context. Hence, they are also implicitly indexed by its fuzzy incidence relation. Sometimes, in particular when we are working with more than one contexts at the same time, we may need to make the index explicit to indicate the context in which the operators are defined. For example, we may have to write the more cumbersome $A^\Box_{\tilde{I}}$ instead of simply $A^\Box$ in such case.

	Given the definition above, we can establish the connection between measures in possibility theory and operators in fuzzy formal context\cite{DuboisSP07}.
	\begin{proposition}\label{prop1}\badstart
		\begin{enumerate}
			\item $A^{+}(m)=\inf_{g\in A}\pi_m(g)=\vartriangle_m\!\!(A)$
			\item $B^{-}(g)=\inf_{m\in B}\pi_g(m)=\vartriangle_g\!\!(B)$
			\item $ A^\Box(m)=\inf_{g\not\in A}1-\pi_m(g)=N_m(A)$
			\item $ B^\Box(g)=\inf_{m\not\in B}1-\pi_g(m)=N_g(B)$
			\item $ A^\Diamond(m)=\sup_{g\in A}\pi_m(g)=\Pi_m(A)$
			\item $ B^\Diamond(g)=\sup_{m\in B}\pi_g(m)=\Pi_g(B)$
		\end{enumerate}
	\end{proposition}
	
As operators in Definition \ref{fuzzyset} map subsets of a domain to fuzzy subsets of its codomain, we can use (strict) $c$-cut to transform them into subsets of the codomain again. Hence, we obtain the following pairs of operators for any $c\in[0,1]$:
	\begin{equation}\label{fop}
		(\cdot)^{+}_{c}: (\mathcal{P}(G), \subseteq)\longleftrightarrow(\mathcal{P}(M),\supseteq):(\cdot)^{-}_{c}
	\end{equation}
	\begin{equation}\label{rop}
		(\cdot)^\Diamond_{>1-c}: (\mathcal{P}(G),\subseteq)\longleftrightarrow(\mathcal{P}(M),\subseteq):(\cdot)^\Box_{c}
	\end{equation}
\begin{theorem}{\rm 
The two pairs of operators in (\ref{fop}) and (\ref{rop}) form Galois connections.}    
\end{theorem}
{\pf} See Appendix A.\epf

This motivates the following definition of cut-based generalizations of concepts.
\begin{definition}
		{\rm For a fuzzy formal context $(G, M, \tilde{I})$, $A\subseteq G$ and $B\subseteq M$ and $c\in [0,1]$, 

              \begin{itemize}
                  \item $(A, B)$  is called a $c$-formal concept if $(A)^{+}_{c}=B$ and $(B)^{-}_{c}=A$
                  \item $(A, B)$  is called a $c$-object oriented concept if  $A^{\Box}_{c}=B$  and $B^{\Diamond}_{>1-c}=A.$
                  \item $(A, B)$ is called a $c$-property oriented concept if $A^{\Diamond}_{>1-c}=B$ and $B^{\Box}_{c}=A.$
              \end{itemize}
        
			}
	\end{definition}
	
	For a fuzzy context $\tilde{\mathbb{K}}$, we denote the set of all $c$-formal concepts,  the set of all $c$-object oriented concepts, and the set of  all $c$-property oriented concepts  by $\mathbf{B}_{c}(\tilde{\mathbb{K}})$, $\mathbf{O}_{c}(\tilde{\mathbb{K}})$, and  $\mathbf{P}_{c}(\tilde{\mathbb{K}})$, respectively.

Each classical formal context $(G, M, I)$ naturally induces a fuzzy  one $(G, M, \tilde{I})$, where the fuzzy relation $\tilde{I}(g, m):=1$ if $(g, m) \in I$ and $0$ if $(g, m) \notin I$. Then, the pairs of maps $((\cdot)^{+}_{1}, (\cdot)^{-}_{1})$, $((\cdot)^{\lozenge}_{>0}, (\cdot)^{\square}_{1})$, and $((\cdot)^{\square}_{1}, (\cdot)^{\lozenge}_{>0})$ are equivalent to the classical pair $(^{+}, ^{-})$, $(\lozenge, \square)$, and $(\square, \lozenge)$, respectively.  As a consequence,  classical  concepts  coincide with the 1-cut concepts if we consider a classical context as a special kind of  fuzzy context. In fact, cut-based concepts enjoy many basic properties of classical ones. Further details regarding this can be found in Appendix A.

\section{Two-sorted Weighted Modal Logic }\label{sec4}
	To represent and reason with information about fuzzy formal context, we present the two-sorted weighted modal logic  {\bf 2WML}. The signature of  {\bf 2WML} consists of two sorts $\{s_1,s_2\}$ and two sets of modalities $\Sigma_{s_1s_2}$ and $\Sigma_{s_2s_1}$ with arities $s_1\ra s_2$ and $s_2\ra s_1$, respectively:
	  \[\Sigma_{s_1s_2}=\{\wbg{c},\wbg{c^+},\wnbg{c}, \wnbg{c^+}\mid c\in[0,1]\},\]	  
	   \[\Sigma_{s_2s_1}=\{\wbm{c},\wbm{c^+},\wnbm{c}, \wnbm{c^+}\mid c\in[0,1]\}.\]  
		Let $P_{s_{1}}$ and $P_{s_{2}}$  be  sets of propositional symbols.  Then,  the  formulas of {\bf 2WML} are  defined by the following grammars, where $\varphi$ and $\psi$ denote formulas of sorts $s_1$ and $s_2$, respectively:
	\[\varphi::=p_{s_1}\mid\neg\varphi\mid\varphi\wedge\varphi\mid\wbm{c}\psi\mid\wbm{c^+}\psi\mid\wnbm{c}\psi\mid\wnbm{c^+}\psi,\]
	\[\psi::=p_{s_{2}}\mid\neg\psi\mid\psi\wedge\psi\mid\wbg{c}\varphi\mid\wbg{c^+}\varphi\mid\wnbg{c}\varphi\mid\wnbg{c^+}\varphi,\]
	where $p_{s_{i}}\in P_{s_{i}}$ and $c\in[0,1]$. The formulas of sorts $s_1$ and $s_2$ are usually called object and property formulas, respectively. We use ${\cal L}^{[\cdot],\wnb{\cdot}}=({\cal L}_{s_1}^{[\cdot],\wnb{\cdot}}, {\cal L}_{s_2}^{[\cdot],\wnb{\cdot}})$ to denote the set of {\bf 2WML} formulas. As usual, we define abbreviations of some common logical connectives and dual modalities as follows (we omit the subscript as the definition applies to both sorts of formulas): $\alpha\vee\beta:=\neg(\neg\alpha\wedge\neg\beta)$,
	$\alpha\ra\beta:=\neg\alpha\vee\beta$, $\alpha\equiv\beta:=(\alpha\ra\beta)\wedge(\beta\ra\alpha)$,  $\langle c\rangle\alpha:=\neg[(1-c)^+]\neg\alpha$, $\langle c^+\rangle\alpha:=\neg[1-c]\neg\alpha$,
	$\langle\!\langle c\rangle\!\rangle\alpha:=\neg[\![(1-c)^+]\!]\neg\alpha$, and $\langle\!\langle c^+\rangle\!\rangle\alpha:=\neg[\![1-c]\!]\neg\alpha$.

    Intuitively, a formula describes a set of objects or attributes in a fuzzy formal context and modal operators correspond to necessity and guaranteed possibility measures. More specifically, propositional symbols are atomic formulas describing basic sets of objects and properties, whereas logical negation and conjunction correspond to set complement and intersection, respectively. On the other hand, modal operators transform formulas of one sort to the other one. They correspond to $c$-cut and strict $c$-cut of operators defined in Definition~\ref{fuzzyset}. Based on the connection established in Proposition~\ref{prop1}, they also express (strict) lower bounds on necessity and guaranteed possibility measures of a set. The semantics below reflects the intuition  precisely.
       
	Formally, we define a {\em fuzzy  context model} as $\mathfrak{M}:=(G, M, \tilde{I},v)$, where $(G,M, \tilde{I})$ is a fuzzy formal context and $v=(v_{1}, v_{2})$ is a  truth valuation function such that $v_{1}:P_{s_{1}}\ra{\mathcal P}(G)$ and $v_{2}:P_{s_{2}}\ra{\mathcal P}(M)$. In the model, $G$ and $M$ are called its $s_1$-sorted and $s_2$-sorted domains, respectively. Below, we define the satisfaction of  formulas in the domains $G$ and  $M$. For atomic, negated, and conjunctive formulas, the definition is standard. The following lists the definition for the satisfaction of modal formulas in $g\in G$ and $m\in M$.
	\begin{itemize}
		\item  $\mathfrak{M},g\models_{s_{1}} \wbm{c}\psi$ iff $N_g(|\psi|)\geq c$
		\item  $\mathfrak{M},g\models_{s_{1}} \wbm{c^+}\psi$ iff $N_g(|\psi|)> c$
		\item  $\mathfrak{M},g\models_{s_{1}} \wnbm{c}\psi$ iff $\vartriangle_g\!\!(|\psi|)\geq c$
		\item  $\mathfrak{M},g\models_{s_{1}} \wnbm{c^+}\psi$ iff $\vartriangle_g\!\!(|\psi|)> c$
		\item  $\mathfrak{M},m\models_{s_{2}} \wbg{c}\varphi$ iff $N_m(|\varphi|)\geq c$
		\item  $\mathfrak{M},m\models_{s_{2}} \wbg{c^+}\varphi$ iff $N_m(|\varphi|)> c$,
		\item  $\mathfrak{M},m\models_{s_{2}} \wnbg{c}\varphi$ iff $\vartriangle_m\!\!(|\varphi|)\geq c$
		\item  $\mathfrak{M},m\models_{s_{2}} \wnbg{c^+}\varphi$ iff $\vartriangle_m\!\!(|\varphi|)> c$.
	\end{itemize}
	where $|\psi|=\{m\in M\mid \mathfrak{M},m\models\psi\}$ and $|\varphi|=\{g\in G\mid \mathfrak{M},g\models\varphi\}$ are truth sets of $\psi$ and $\varphi$, respectively and the necessity and guaranteed possibility measures are induced from respective possibility distributions $\pi_g$ for $g\in G$ and $\pi_m$  for $m\in M$. 
     
	To further see the intuition behind the semantics of modal formulas in {\bf 2WML}, let us consider the case of $\wbm{c}\psi$. Expanding the definition of the necessity measure based on the possibility distribution $\pi_g$, we have
	$N_g(|\psi|)=\bigwedge_{m\in M}(\tilde{I}(g,m)\ra\psi(m))$,
	where $\psi(m)=1$ if $\mathfrak{M},m\models_{s_{1}}\psi$ and $\psi(m)=0$ elsewhere. Hence, $N_g(|\psi|)$ specifies the degree that for any attribute $m$ related to $g$, $m$ satisfies $\psi$. Thus,  $\mathfrak{M},g\models_{s_{2}}\wbm{c}\psi$ means that such a degree is at least $c$.  Analogously, because
	$\vartriangle_g\!\!(|\psi|)=\bigwedge_{m\in M}(\psi(m)\ra\tilde{I}(g,m))$
	specifies the degree that for any attribute $m$ satisfying $\psi$,  $m$ is related to $g$, $\mathfrak{M},g\models_{s_{2}}\wnbm{c^+}\psi$ means that such a degree is greater than $c$. 
	\begin{myex}\label{ex2}{\rm 
		Returning to Example~\ref{ex1}, let us slightly modify the context. Assume that $G$ and $M$ remain the same, and based on the users' download history, we can infer whether a user is interested in a document. However, such inference cannot be precise by its nature. Hence, there is some kind of uncertainty about a user's real interest in a document, and we can use $\tilde{I}$ to denote such uncertain information. More specifically, $\tilde{I}(g,m)$ denotes the degree of possibility of $g$ being interested in $m$. Let $\varphi$ and $\psi$ be formulas describing the sets $A$ and $B$, respectively. Then, for a fuzzy context model $\mathfrak{M}$ in which $|\varphi|=A$ and $|\psi|=B$, we have the following interpretation  of formulas:
        \begin{itemize}
			\item $\mathfrak{M},g\models [0.9]_{\mathbf p}\psi$:  it is highly certain that $g$ is only interested in documents in $B$
			\item $\mathfrak{M},m\models [0.9]_{\mathbf o}\varphi$: it is highly certain that only readers in $A$ are interested in the document $m$
			\item $\mathfrak{M},g\models\wnbm{0.9}\psi$: it is highly certain that the reader $g$ is interested in all documents  in $B$
			\item $\mathfrak{M},m\models\wnbg{0.9}\varphi$:  it is highly certain that all readers in $A$  are interested in the document $m$
		\end{itemize}}
	\end{myex}

\begin{myex}\label{ex3}{\rm 
		Continuing with Example~\ref{ex2}, let us consider an alternative interpretation of the fuzzy binary relation. Assume that, through users' feedback, we have information about users' real interests. Hence, $\tilde{I}$ can denote how strongly a reader is interested in a document. In other words, $\tilde{I}(g,m)$ is the degree of strength of $g$ being interested in $m$.  Then, in the  fuzzy context model $\mathfrak{M}$, the interpretation of formulas in the previous example becomes
        \begin{itemize}
			\item $\mathfrak{M},g\models [0.9]_{\mathbf p}\psi$:  $g$ is hardly interested in documents outside $B$
			\item $\mathfrak{M},m\models [0.9]_{\mathbf o}\varphi$: readers outside $A$ are hardly interested in the document $m$
			\item $\mathfrak{M},g\models\wnbm{0.9}\psi$:   $g$ is highly interested in all documents  in $B$
			\item $\mathfrak{M},m\models\wnbg{0.9}\varphi$: all readers in $A$ are  highly interested in the document $m$
		\end{itemize}}
	\end{myex}

Given the satisfaction of modal formulas and Proposition~\ref{prop1}, we can establish the connection between the semantics of {\bf 2WML} and operators in fuzzy formal contexts.
	
	\begin{proposition}\label{prop2}{\rm For any object formula $\varphi$ and property formula $\psi$, we have
\[\begin{array}{llll}
1.~|\varphi|^{+}_{>c}=|\wnbg{c^+}\varphi|,\;& 2.~|\varphi|^{+}_{c}=|\wnbg{c}\varphi|,\;& 3.~|\varphi|^{\square}_{>c}=|\wbg{c^+}\varphi|,\;& 4.~ |\varphi|^{\square}_{c}=|\wbg{c}\varphi|,\\
5.~|\psi|^{-}_{>c}=|\wnbm{c^+}\psi|,& 6.~|\psi|^{-}_{c}=|\wnbm{c}\psi|,& 7.~|\psi|^{\square}_{>c}=|\wbm{c^+}\psi|,& 8.~|\psi|^{\square}_{c}=|\wbm{c}\psi|.
\end{array}\]        
}
\end{proposition}
	As a corollary, we also have the connection between the derived modalities $\langle c\rangle$ and $\langle c^+\rangle$ and upper approximation operators as follows.
	\begin{mycor}{\rm For any object formula $\varphi$ and property formula $\psi$, we have\\
		\begin{inparaenum}
			\item $|\varphi|^{\lozenge}_{>c}=|\langle c^+\rangle_\mathbf{o}\varphi|$,
			\item $|\varphi|^{\lozenge}_{c}=|\langle c\rangle_\mathbf{o}\varphi|$,
			\item $|\psi|^{\lozenge}_{>c}=|\langle c^+\rangle_\mathbf{p}\psi|$,
			\item $|\psi|^{\lozenge}_{c}=|\langle c\rangle_\mathbf{p}\psi|$.
		\end{inparaenum}}
	\end{mycor}
Next, we define some basic notions in the semantics of {\bf 2WML}.
\begin{definition}\label{semnotions}{\rm
Let $\Gamma$ be a set of $s$-sorted formulas and let $\mathfrak{M}$ be a model. Then, for any element $w$ in its $s$-sorted domain, $\mathfrak{M},w\models_s\Gamma$ if $\mathfrak{M},w\models_s\phi$ for all $\phi\in\Gamma$.  In addition, $\Gamma$ is {\em satisfiable} if there exist a model $\mathfrak{M}$ and a $w$ in its $s$-sorted domain such that $\mathfrak{M},w\models_s\Gamma$. We say that $\Gamma$ is {\em finitely satisfiable} if every finite subset of $\Gamma$ is satisfiable.

Let $\mathbf{C}$ be a class of models. Then, for a set $\Gamma\cup\{\phi\}$ of $s$-sorted formulas, $\phi$ is a {\em local semantic consequence\/} of $\Gamma$ over $\mathbf{C}$ and denoted as $\Gamma\models^{\mathbf{C}}_{s}\phi$  if $\mathfrak{M}, w\models_{s}\Gamma$ implies $\mathfrak{M}, w\models_{s}\phi$ for all models $\mathfrak{M}\in\mathbf{C}$ and $w$ in its $s$-sorted domain. If $\mathbf{C}$ is the class of all models, we omit the superscript and denote it by $\Gamma\models_{s}\phi$. When $\Gamma=\emptyset$, we say that $\phi$ is valid and simply write  $\models^{\mathbf{C}}_{s}\phi$ or $\models_{s}\phi$. Note that, for a finite set of formulas $\Gamma$, $\Gamma\models_{s}\phi$ is equivalent to the validity of $\bigwedge\Gamma\rightarrow\phi$.}
\end{definition}

\begin{figure}[h]
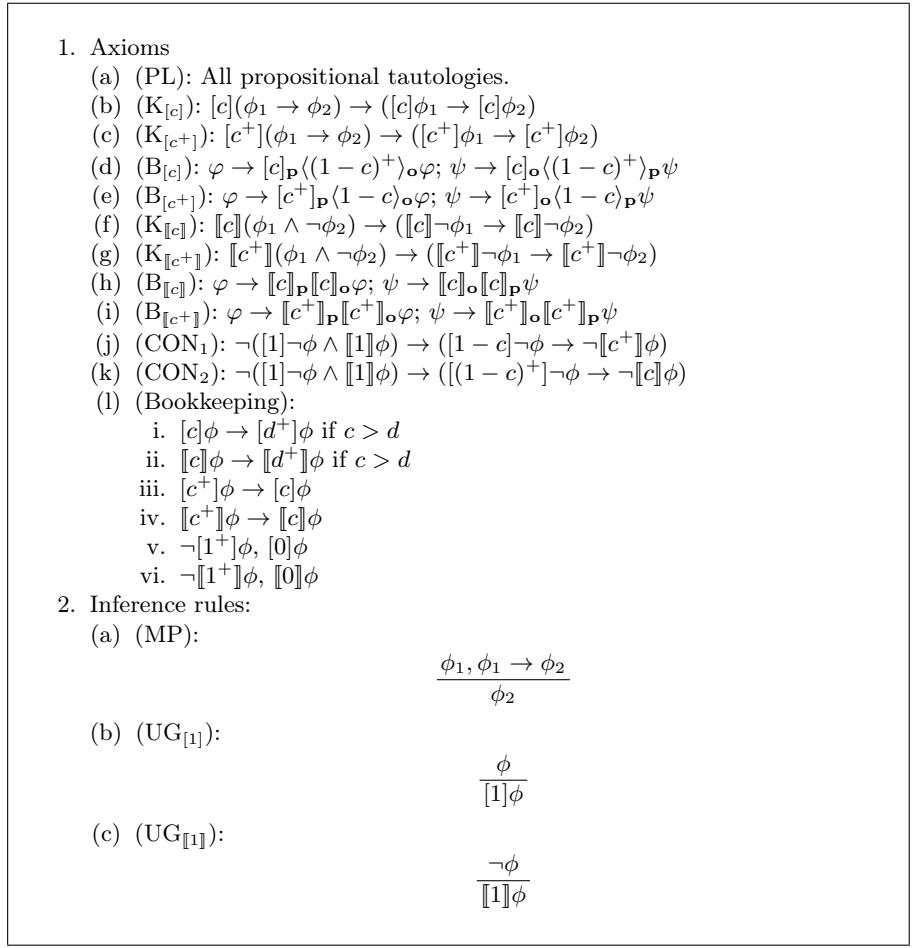
\centering
		\framebox[120mm] {\parbox{110mm}{\begin{enumerate}
					\item Axioms
					\begin{enumerate}
						\item (PL): All propositional tautologies.
						\item (K$_{[c]}$): $[c](\phi_{1}\rightarrow\phi_{2})\rightarrow([c]\phi_{1}\rightarrow[c]\phi_{2})$
						\item (K$_{[c^{+}]}$): $[c^{+}](\phi_{1}\rightarrow\phi_{2})\rightarrow([c^{+}]\phi_{1}\rightarrow[c^{+}]\phi_{2})$ 
						\item (B$_{[c]}$): $\varphi\rightarrow [c]_{\mathbf p} \langle (1-c)^{+}\rangle_{\mathbf o} \varphi$; $\psi\rightarrow [c]_{\mathbf o} \langle(1-c)^{+}\rangle_{\mathbf p} \psi$ 
						\item (B$_{[c^+]}$): $\varphi\rightarrow [c^{+}]_{\mathbf p}\langle 1-c\rangle_{\mathbf o} \varphi$; $\psi\rightarrow[c^+]_{\mathbf o} \langle 1-c\rangle_{\mathbf p} \psi$  
						\item (K$_{[\![ c]\!] }$): $[\![ c]\!] (\phi_{1}\wedge\neg \phi_{2})\rightarrow ([\![ c]\!] \neg\phi_{1}\rightarrow [\![ c]\!] \neg\phi_{2})$ 
                        \item (K$_{[\![ c^{+}]\!] }$): $[\![ c^{+}]\!] (\phi_{1}\wedge\neg \phi_{2})\rightarrow ([\![ c^{+}]\!] \neg\phi_{1}\rightarrow [\![ c^{+}]\!] \neg\phi_{2})$ 
						\item (B$_{[\![c]\!] }$): $ \varphi\rightarrow \wnbm{c}\wnbg{c}\varphi$; $ \psi\rightarrow \wnbg{c}\wnbm{c}\psi$ 
						\item (B$_{[\![c^{+}]\!] }$): $ \varphi\rightarrow \wnbm{c^+}\wnbg{c^+}\varphi$; $ \psi\rightarrow \wnbg{c^+}\wnbm{c^+}\psi$ 
                        \item (CON$_1$): $\neg([1]\neg\phi\wedge\wnb{1}\phi)\ra([1-c] \neg\phi\rightarrow\neg[\![c^{+}]\!] \phi)$
						\item (CON$_2$): $\neg([1]\neg\phi\wedge\wnb{1}\phi)\ra([(1-c)^{+}] \neg\phi\rightarrow\neg[\![c]\!] \phi)$
                        \item (Bookkeeping):
                        \begin{enumerate}
                        \item $[c]\phi\rightarrow[d^+]\phi$ if $c>d$
                        \item $\wnb{c}\phi\rightarrow\wnb{d^+}\phi$ if $c>d$
                        \item $[c^+]\phi\rightarrow[c]\phi$
                        \item $\wnb{c^+}\phi\rightarrow\wnb{c}\phi$ 
                        \item $\neg[1^+]\phi$, $[0]\phi$
                        \item $\neg\wnb{1^+}\phi$,  $\wnb{0}\phi$   
                        \end{enumerate}
					\end{enumerate}
					\item  Inference rules: 
					\begin{enumerate}
						\item (MP): \[\infer{\phi_2}{\phi_1, \phi_1\rightarrow \phi_2}\]
						\item (UG$_{[1] })$:
                        \[\infer{[1] \phi}{\phi}\]
						\item (UG$_{[\![1]\!] }$): 
						\[\infer{[\![1]\!] \phi}{\neg\phi}\]
					\end{enumerate}
				\end{enumerate}
		}} \caption{The axiomatic system $\mathbf{2WML}$ \label{KB}}
	\end{figure}

To axiomatize the local semantic consequence in  \textbf{2WML}, we present a Hilbert-style axiomatic system in Figure~\ref{KB}. In the system, we omit the subscript  $\mathbf o$ or $\mathbf p$ in modal operators as the axioms and rules hold for both sorts of formulas, and we use $\phi$, $\phi_1$, and $\phi_2$ to denote formulas when we do not specify their sorts particularly. By contrast, in axioms (d), (e), (h), and (i), $\varphi$ and $\psi$ are formulas of sorts $s_1$ and  $s_2$, respectively.
The axioms (K$_{[c]}$) and (K$_{[c^+]}$) are the weighted analogue of the standard (K) axiom in normal modal logic systems. The axioms (B$_{[c]}$) and (B$_{[c^+]}$) stipulates that the two possibility distributions $\pi_g$ and $\pi_m$ are mutually symmetric in the sense that $\pi_g(m)=\pi_m(g)$ for any $g\in G$ and $m\in M$. They are also the weighted analogue of the (B) axiom in normal modal logic for symmetric Kripke frames and converse axiom in temporal logic. Axioms (K$_{\wnb{c}}$),(K$_{\wnb{c^+}}$), (B$_{\wnb{c}}$), and (B$_{\wnb{c^+}}$) correspond to the translation of necessity modalities to sufficiency ones in a complemented context, which will be detailed in the next section. The (CON$_1$) and (CON$_2$) axioms describe the connection between necessity and guaranteed possibility measures. They rely on the fact that $\Pi(X)\geq\vartriangle\!\!(X)$ for any nonempty subset $X$ of a domain. The antecedent of these two axioms $\neg([1]\neg\phi\wedge\wnb{1}\phi)$ simply excludes the exceptional case when $\Pi(\emptyset)=0$ and $\vartriangle\!\!(\emptyset)=1$. Finally, the (Bookkeeping) axioms specify the boundaries of the weights and their ordereding. The inference rules include the standard necessitation rule and its translation for sufficiency modalities.

\begin{definition}{\rm 
In the axiomatic system $\mathbf{2WML}$, a sequence of formulas $\phi_{1}, \phi_{2}, \ldots\phi_{n}$ is called a {\em proof\/} for the formula $\phi$  if  $\phi_{n}=\phi$ and  $\phi_{i}$ is an instance of its axioms or inferred from $\phi_{1}, \ldots, \phi_{i-1}$ using modus pones (MP) or universal generalization (UG) rules.  If an $s$-sorted formula $\phi$  has a proof in $\mathbf{2WML}$,  we say that $\phi$ is a {\em theorem\/} and write $\vdash_{s}\phi$.
Let $\Gamma\cup\{\phi\}$ be a set of $s$-sorted formulas. Then, we say that $\phi$ is {\em provable\/} (or derivable) form $\Gamma$, denoted by $\Gamma\vdash_{s}\phi$, if there exist $\phi_1,\ldots,\phi_n\in\Gamma$ such that $\vdash_{s}(\phi_1\wedge\ldots\wedge\phi_n)\rightarrow\phi$. In addition, the set $\Gamma$ is inconsistent  if $\bot$ is provable from the set, otherwise it is consistent.}
\end{definition}

\begin{mylma}
		{\rm The following are derivable in the system {\bf 2WML}\begin{enumerate}
				\item $\neg([1]\neg\phi\wedge\wnb{1}\phi)\ra(\neg\langle c^{+}\rangle\phi\rightarrow\langle\!\langle 1-c\rangle\!\rangle\neg\phi)$.
				\item $\neg([1]\neg\phi\wedge\wnb{1}\phi)\ra(\neg\langle c\rangle\phi\rightarrow\langle\!\langle (1-c)^{+}\rangle\!\rangle\neg\phi)$.
		\end{enumerate}}
	\end{mylma}
{\pf} It follows from axioms CON$_1$ and CON$_2$ directly. \epf
    	
\begin{theorem}
	    {\rm The system {\bf 2WML} is sound with respect to the class of all fuzzy context models. That is, for any $s\in \{s_{1}, s_{2}\}$, $\Gamma\vdash_{s}\phi$ implies that $\Gamma\models_{s}\phi$ for any set of $s$-sorted formulas $\Gamma\cup\{\phi\}$.}
	\end{theorem}

\subsection{Application}	
As an application of our logic, we show that it can represent different notions of $c$-cut concepts in fuzzy formal contexts.

\begin{definition}
    {\rm Let $\mathbf{C}$ denote the  class of all fuzzy context models based on a  fuzzy context $\mathbb{K}$. Then, for $c\in[0,1]$, we define $Fm^{c}_{PC}, Fm^{c}_{OC}\subseteq {\mathcal L}^{[\cdot]}_{s_{1}}\times{\mathcal L}^{[\cdot]}_{s_2}$ by
\begin{itemize}
\item[(a)] $Fm^{c}_{PC}:=\{(\varphi, \psi) \mid \models^\mathbf{C}_{s_{1}} \varphi\equiv[c]_{\mathbf p}\psi, \models^\mathbf{C}_{s_{2}}  \langle(1-c)^{+}\rangle_{\mathbf o}\varphi\equiv\psi\}$
 \item[(b)] $Fm^{c}_{OC}:=\{(\varphi, \psi) \mid    \models^\mathbf{C}_{s_{1}} \varphi\equiv \langle(1-c)^{+}\rangle_{\mathbf p}\psi,  \models^\mathbf{C}_{s_{2}} [c]_{\mathbf o}\varphi\equiv\psi\}$
\item[(c)] $Fm^{c}_{PC_{ext}}:=\pi_1(Fm^{c}_{PC})$ and $Fm^{c}_{PC_{int}}:=\pi_2(Fm^{c}_{PC})$
\item[(d)] $Fm^{c}_{OC_{ext}}:=\pi_1(Fm^{c}_{OC})$ and $Fm^{c}_{OC_{int}}:=\pi_2(Fm^{c}_{OC})$
    \end{itemize}
where $\pi_1$ and $\pi_2$ are projection operators\footnote{That is, for a subset $S\subseteq A\times B$, $\pi_1(S)=\{a\in A\mid \exists b\in B, (a.b)\in S\}$ and $\pi_2(S)=\{b\in B\mid \exists a\in A, (a.b)\in S\}$}.}
\end{definition}
By the definition, we have
$Fm^c_{PC_{ext}}=\{\varphi\in {\cal L}^{[\cdot]}_{s_1}\mid~ \models^\mathbf{C}_{s_{1} } [c]_{\mathbf p}\langle(1-c)^{+}\rangle_{\mathbf o}\varphi\equiv\varphi\}$, $Fm^c_{PC_{int}}=\{\varphi\in  {\cal L}^{[\cdot]}_{s_2}\mid~ \models^\mathbf{C}_{s_{2} } \langle(1-c)^{+}\rangle_{\mathbf o}[c]_{\mathbf p}\varphi\equiv\varphi\}$, $Fm^c_{OC_{ext}}=\{\varphi\in  {\cal L}^{[\cdot]}_{s_{1}}\mid ~ \models^\mathbf{C}_{s_{1} } \langle(1-c)^{+}\rangle_{\mathbf p}[c]_{\mathbf o}\varphi\equiv\varphi\}$, and $Fm^c_{OC_{int}}=\{\varphi\in  {\cal L}^{[\cdot]}_{s_{2}}\mid ~ \models^\mathbf{C}_{s_{2} }[c]_{\mathbf o}\langle(1-c)^{+}\rangle_{\mathbf p}\varphi\equiv\varphi\}$.

Note that these sets are implicitly parameterized by the underlying fuzzy context and should be indexed with $\mathbb{K}$. However, for simplicity, we omit the index. Obviously, when $(\varphi, \psi)\in Fm^{c}_{PC}$, we have $(|\varphi|, |\psi|)\in\mathbf{P}_{c}(\tilde{\mathbb{K}})$ for any $\tilde{\mathbb{K}}$-based models. Hence, $Fm^{c}_{PC}$ consists of pairs of formulas representing $c$-property oriented concepts. Analogously, $Fm^{c}_{OC}$ provides the representation of $c$-object oriented concepts. 

\begin{definition}
    {\rm Let $\tilde{\mathbb{K}}=(G,M,\tilde{I})$ be a fuzzy formal context and let $\mathbf{C}$ be the class of all $\tilde{\mathbb{K}}$=based models. Then, we define $Fm^{c}_{FC}\subseteq {\cal L}^{\wnb{\cdot}}_{s_1}\times {\cal L}^{\wnb{\cdot}}_{s_2}$ by 
    \begin{itemize}
     \item[(a)] $Fm^{c}_{FC}:=\{(\varphi, \psi) \mid   \models^\mathbf{C}_{s_1} \varphi\equiv [\![c]\!]_{\mathbf p}\psi,  \models^\mathbf{C}_{s_2}  [\![c]\!]_{\mathbf o}\varphi\equiv \psi\}$
    \item[(b)] $Fm^{c}_{FC_{ext}}:=\pi_1(Fm^{c}_{FC})$ and $Fm^{c}_{FC_{int}}:=\pi_2(Fm^{c}_{FC})$
    \end{itemize}}
\end{definition}
From the definition, we can derive $Fm^{c}_{FC_{ext}}=\{\varphi\in {\cal L}^{\wnb{\cdot}}_{s_1}\mid ~\models^\mathbf{C}_{s_1} [\![c]\!]_{\mathbf p}[\![c]\!]_{\mathbf o}\varphi\equiv\varphi\}$ and $Fm^{c}_{FC_{int}}=\{\varphi\in {\cal L}^{\wnb{\cdot}}_{s_{2}}\mid ~ \models^\mathbf{C}_{s_{2} } [\![c]\!]_{\mathbf o}[\![c]\!]_{\mathbf p}\varphi\equiv\varphi\}$. 
Hence, the set $Fm^{c}_{FC}$ provides a logical representation of  $c$-formal concepts induced from the fuzzy context $(G, M, \tilde{I})$.

\begin{myex}\label{ex4}{\rm 
		Continuing with Example~\ref{ex3}, if $(\varphi,\psi)\in Fm_{FC}^{0.9}$, then according to the interpretation given in that example, $|\varphi|$ is the set of users who are highly interested in all documents in $|\psi|$, and $|\psi|$ is the set of documents that strongly attract common interest of all users in $|\varphi|$. This means that $\varphi$ and $\psi$ describe a group of users and a type of documents that have strong connection. For instance, $\varphi$ and $\psi$ may denote a reading club for a particular author and  all books written by the author, respectively.
        }        
\end{myex}	
\section{Necessity and Sufficiency Fragments of {\bf 2WML}}\label{sec5}
While we have established the soundness of {\bf 2WML} with respect to fuzzy context models, it remains an open question whether the logic is also complete. However, we can show that the necessity and sufficiency fragments of {\bf 2WML} are both sound and complete with respect to fuzzy context models. The language of the necessity fragment includes all propositional connectives along with the modal operators $[c]$ and $[c^{+}]$, whereas the language of the sufficiency fragment is formed using the modalities $[\![c]\!]$ and 
$[\![c^{+}]\!]$.

\subsection{The system {\bf 2WKB}}
In this section, we discuss the necessity fragment of {\bf 2WML}. The logic is called {\bf 2WKB}. Its signature is derived from that of {\bf 2WML} omitting sufficiency modalities $\wnb{c}$ and $\wnb{c^+}$. Syntax and semantics, particularly the definition of formulas and their satisfaction, are inherited from {\bf 2WML} without modification.  We use ${\cal L}^{[\cdot]}=({\cal L}_{s_1}^{[\cdot]}, {\cal L}_{s_2}^{[\cdot]})$ to denote the set of {\bf 2WKB} formulas.  The corresponding axiomatic system for {\bf 2WKB} consists of axioms (a)-(e), (l)(i,iii,v), and inference rules (a) and (b) of {\bf 2WML}. Also, we still use $\models$ and $\vdash$ to denote the semantic consequence and provability relations in {\bf 2WKB}, respectively. Essentially, {\bf 2WKB} is a weighted extension of the two-sorted modal logic {\bf KB} introduced in \cite{ijcrs23,bmlpl} and a two-sorted variant of the quantitative modal logic (QML) introduced in \cite{qml1}. Hence, we can prove the completeness of {\bf 2WKB} employing a synergy of the canonical model construction techniques used in {\bf KB} and QML. 

As in QML,  {\bf 2WKB} lacks the compactness theorem in its semantics. That is, a finitely satisfiable set of formulas is not necessarily satisfiable in  {\bf 2WKB}. However, since the derivation in an axiomatic system is always finitary, we can not have the completeness without any restriction. Hence, we restrict our consideration to the case that only finite number of weights is allowed in the set of formulas. For this, we define the set of degrees occurring (explicitly or implicitly) in a formula $\phi$ as
\[deg(\phi)=\{0,1\}\cup\{c\mid [c]_i, [1-c]_i, [c^+]_i,~\mbox{\rm or}~[(1-c)^+]_i~\mbox{\rm appears in}~\phi~ \mbox{\rm for some}~i={\mathbf o}, {\mathbf p} \}, \]
and for a set of formulas $\Phi$, let $deg(\Phi):=\bigcup_{\phi\in\Phi}deg(\phi)$. Moreover, for any subset $D\subseteq[0,1]$, we use ${\cal L}(D)=({\cal L}_{s_1}(D),{\cal L}_{s_2}(D))$ to denote the set of  {\bf 2WKB} formulas in which only degrees in $D$ occur (for simplicity, we omit the superscript $[\cdot]$ here).

We can now state the main theorem of this section.
\begin{theorem}
	   \label{soucomkbwml} {\rm {\bf 2WKB} is sound and complete with respect to the class of all fuzzy context models. That is for any $s\in \{s_{1}, s_{2}\}$, $\Gamma\vdash_{s}\phi$ iff $\Gamma\models_{s}\phi$ where $\Gamma\cup\{\phi\}$ is a set of $s$-sorted formulas such that $deg(\Gamma\cup\{\phi\})$ is finite.}
\end{theorem}
{\pf} The soundness follows from that of {\bf 2WML}. Here, we simply sketch the proof of the completeness and its details are included in Appendix B. Suppose that $\Gamma\not\vdash_{s}\phi$. Hence, $\Gamma\cup\{\neg\phi\}$ is consistent. By the standard Lindenbaum lemma, $\Gamma\cup\{\neg\phi\}$ can be extended to a maximally consistent subset $\Phi_s$ of $s$-sorted  formulas. Then, by the canonical model construction, we construct a model whose domains are the sets of all maximally consistent subsets. The model satisfies the truth lemma, that is, a formula is true in an element of the domains iff it belongs to that element. Thus, we have a model and its domain element $\Phi_s$ in which all formulas in $\Gamma$ is true but $\phi$ is false, i.e. $\Gamma\not\models_{s}\phi$.\epf 

To construct the canonical model, let us consider the set of two-sorted formulas ${\cal L}(D)=({\cal L}_{s_1}(D),{\cal L}_{s_2}(D))$, where $D=deg(\Gamma\cup\{\phi\})$. For any $\Sigma_g\subseteq {\cal L}_{s_1}(D)$, $\Sigma_m\subseteq {\cal L}_{s_2}(D)$, and $c\in D$, we define the following notations.
\[\Sigma_g/[c]_{\mathbf p}=\{\psi\mid d\geq c, [d]_{\mathbf p}\psi\in\Sigma_g~\mbox{\rm or}~[d^+]_{\mathbf p}\psi\in\Sigma_g\}\]
\[\Sigma_m/[c]_{\mathbf o}=\{\varphi\mid d\geq c, [d]_{\mathbf o}\varphi\in\Sigma_m~\mbox{\rm or}~[d^+]_{\mathbf o}\varphi\in\Sigma_m\}\]
\[\Sigma_g/[c^+]_{\mathbf p}=\{\psi\mid d>c, [d]_{\mathbf p}\psi\in\Sigma_g\} 
\cup\{\psi\mid d\geq c, [d^+]_{\mathbf p}\psi\in\Sigma_g\}.\]
\[\Sigma_m/[c^+]_{\mathbf o}=\{\varphi\mid d>c, [d]_{\mathbf o}\varphi\in\Sigma_g\} 
\cup\{\varphi\mid d\geq c, [d^+]_{\mathbf o}\varphi\in\Sigma_g\}.\]

Based on these notations, we have the following lemma.
\begin{mylma}\label{bcondition}
{\rm Let  $\Sigma_g$  and $\Sigma_m$ be maximally consistent subsets of ${\cal L}_{s_1}(D)$ and ${\cal L}_{s_2}(D)$, respectively. Then, we have
\begin{enumerate}
    \item $\Sigma_g/[c]_{\mathbf p}\subseteq\Sigma_m$ iff $\Sigma_m/[c]_{\mathbf o}\subseteq\Sigma_g$,
    \item $\Sigma_g/[c^+]_{\mathbf p}\subseteq\Sigma_m$ iff $\Sigma_m/[c^+]_{\mathbf o}\subseteq\Sigma_g$.
\end{enumerate}}
\end{mylma}
With this lemma, we can now define the notion of canonical models.
\begin{definition}\label{canmodel}{\rm A fuzzy context model ${\mathfrak M}=(G,M,\tilde{I},v)$ is called a ${\mathcal L}(D)$-canonical model if \begin{itemize}
    \item $G$ is the set of all maximally consistent subsets of ${\mathcal L}_{s_1}(D)$
    \item $M$ is the set of all maximally consistent subsets of ${\mathcal L}_{s_2}(D)$
    \item $\tilde{I}$ satisfies the following two conditions for all $\Sigma_g\in G$ and $\Sigma_m\in M$:
    \begin{itemize}
        \item if $c>0$, then $\Sigma_g/[c]_{\mathbf p}\subseteq\Sigma_m$ (or equivalently $\Sigma_m/[c]_{\mathbf o}\subseteq\Sigma_g$) iff $\tilde{I}(\Sigma_g,\Sigma_m)>1-c$
        \item $\Sigma_g/[c^+]_{\mathbf p}\subseteq\Sigma_m$ (or equivalently $\Sigma_m/[c^+]_{\mathbf o}\subseteq\Sigma_g$) iff $\tilde{I}(\Sigma_g,\Sigma_m)\geq 1-c$
    \end{itemize}
    \item the valuation $v=(v_1,v_2)$ is defined by $v_1(p_{s_1})=\{\Sigma_g\mid p_{s_1}\in\Sigma_g\}$ and $v_2(p_{s_2})=\{\Sigma_m\mid p_{s_2}\in\Sigma_m\}$. 
\end{itemize}
}
\end{definition}
The conditions on the fuzzy incidence relation above only specify a range of possible values between elements of $G$ and $M$ but do not give us a unique canonical model constructively. The following lemma shows that canonical models indeed exist.
\begin{mylma}{\rm [Model Existence Lemma]
There exists a fuzzy binary relation $\tilde{I}:G\times M\rightarrow[0,1]$ satisfying the two conditions for the definition of canonical models.   
}\end{mylma}
Finally, we finish the proof of the completeness by providing the truth lemma.
\begin{mylma}{\rm [Truth Lemma] 
Let  ${\mathfrak M}=(G,M,\tilde{I},v)$ be a ${\mathcal L}(D)$-canonical model and let $\varphi\in{\mathcal L}_{s_1}(D)$ and $\psi\in{\mathcal L}_{s_2}(D)$. Then, for any $\Sigma_g\in G$ and $\Sigma_m\in M$, we have
\begin{enumerate}
    \item ${\mathfrak M},\Sigma_g\models_{s_1}\varphi$ iff $\varphi\in\Sigma_g$ and
    \item ${\mathfrak M},\Sigma_m\models_{s_2}\psi$ iff $\psi\in\Sigma_m$. 
\end{enumerate}}
\end{mylma}
\subsection{The system {\bf 2WKF}}
This section is devoted to the sufficiency fragment of {\bf 2WML}, called {\bf 2WKF}. Its signature is obtained by removing the modalities $[c]$ and $[c^{+}]$
from the full language of {\bf 2WML}. Despite this reduction, its syntax and semantics, particularly the notions of formula formation and satisfaction remain the same with those defined for {\bf 2WML}.  We use ${\cal L}^{\wnb{\cdot}}=({\cal L}_{s_1}^{\wnb{\cdot}}, {\cal L}_{s_2}^{\wnb{\cdot}})$ to denote the set of {\bf 2WKF} formulas.  The axiomatization for {\bf 2WKF} consists of axioms (a), (f)-(i), and (l) (ii, iv, vi), and rules (a) and (c) in the axiomatic system of {\bf 2WML}.  Also, {\bf 2WKF} is the weighted extension of the two-sorted modal logic {\bf KF} introduced in \cite{ijcrs23,bmlpl}. Hence, as in the case of {\bf KF}, we prove the completeness of {\bf 2WKF} by a translation mapping between {\bf 2WKF} and {\bf 2WKB}. As the  completeness of {\bf 2WKF} depends on that of  {\bf 2WKB}, they have the same restriction. For this, we define $deg(\phi)$ for a  {\bf 2WKF} formula $\phi$ and extend it to  a set of formulas as in the case of  {\bf 2WKB} except that necessity modalities are replaced with sufficiency ones.   
\begin{theorem}
	    \label{soucomkfwml}{\rm {\bf 2WKF} is sound and complete with respect to the class of all fuzzy context models. That is, for any $s\in \{s_{1}, s_{2}\}$, $\Gamma\vdash_s\phi$ iff $\Gamma\models_{s}\phi$ where $\Gamma\cup\{\phi\}$ is a set of $s$-sorted formulas such that $deg(\Gamma\cup\{\phi\})$ is finite.}
	\end{theorem}

To establish this result, we first define a translation mapping $\rho=(\rho_{1}, \rho_{2})$ that maps each formula in {\bf 2WKF} to one of the same sort in {\bf 2WKB}  as follows:
\begin{enumerate}
\item $\rho_{i}(p):=p$ for all $p\in P_{s_{i}}$ where $i=1,2$,
\item $\rho_{i}(\phi_1\wedge\phi_2):=\rho_{i}(\phi_1)\wedge\rho_{i}(\phi_2)$,
\item $\rho_{i}(\neg\phi):=\neg\rho_{i}(\phi)$,
\item $\rho_{1}([\![x]\!]_{\mathbf p}\psi):=[x]_{\mathbf p}\neg\rho_{2}(\psi)$ for $s_2$-sorted formula $\psi$  and $x\in \{c,c^{+}\}$,
\item $\rho_{2}([\![x]\!]_{\mathbf o}\varphi):=[x]_{\mathbf o}\neg\rho_{1}(\varphi)$ for $s_1$-sorted formula $\varphi$ and $x\in \{c, c^{+}\}$,
\end{enumerate}
where $\phi, \phi_1, \phi_2$ are $s_i$-sorted formulas for $i=1,2$. Also, for a set $\Gamma$ of \textbf{2WKF} formulas, we write $\rho(\Gamma):=\{\rho(\phi)\mid \phi\in\Gamma\}$.

An inverse translation  mapping $\rho^{-1}$ can also be similarly defined.  The only difference is that we exchange the position of $\wnb{x}$ and $[x]$ in items 4 and 5 above. By induction on the complexity of formulas, we can derive $\vdash_s\phi\equiv\rho^{-1}(\rho(\phi))$ and $\vdash_s\phi\equiv\rho(\rho^{-1}(\phi))$ in {\bf 2WKF} and {\bf 2WKB}, respectively.

Using these two mapping, we can prove the following correspondence theorem. For simplicity, we use the same notations $\vdash$ and $\models$ for both logics without confusion, as the formulas to which the relations are applied can disambiguate them.

\begin{theorem}
    \label{translation}
          {\rm  Let $\Gamma\cup\{\phi\}$ be a set of $s$-sorted \textbf{2WKF} formulas.  Then,
          \begin{enumerate}
              \item $\Gamma\vdash_s\phi$ iff $\rho(\Gamma)\vdash_s\rho(\phi)$
              \item Let $\mathfrak{M}:=(G,M,\tilde{I},v)$ be a fuzzy context model and let $\overline{\mathfrak{M}}:=(G,M,\tilde{J}, v)$ be its complemented model defined by $\tilde{J}(g,m)=1-\tilde{I}(g,m)$ for any $g\in G$ and $m\in M$. Then, for any  $g\in G, \mathfrak{M}, g\models_{s_1} \phi$ iff $\overline{\mathfrak{M}}, g\models_{s_1} \rho_1(\phi)$, and  for any  $m\in M, \mathfrak{M}, m\models_{s_2} \phi$ iff $\overline{\mathfrak{M}}, m\models_{s_2} \rho_2(\phi)$
              \item $\Gamma\models_s\phi$  iff $\rho(\Gamma)\models_s\rho(\phi)$ .
          \end{enumerate}
           }
\end{theorem}
{\pf} \begin{enumerate}
    \item By the definition of provability relation, we only need to prove $\vdash_s\phi$ iff $\vdash_s\rho(\phi)$. For the forward direction, if $\phi_1,\phi_2,\cdots,\phi_k=\phi$ is a proof of $\phi$ in {\bf 2WKF}, then $\rho(\phi_1),\rho(\phi_2),\cdots,\rho(\phi_k)=\rho(\phi)$ is a proof of $\rho(\phi)$ in {\bf 2WKB} since $\rho$ transforms each instance of axioms or rules in {\bf 2WKF} to one in {\bf 2WKB}. For the backward direction, analogously, if $\rho(\phi_1),\rho(\phi_2),\cdots,\rho(\phi_k)=\rho(\phi)$ is a proof of $\rho(\phi)$ in {\bf 2WKB}, then  $\rho^{-1}(\rho(\phi_1)),\rho^{-1}(\rho(\phi_2)),\cdots,\rho^{-1}(\rho(\phi_k))=\rho^{-1}(\rho(\phi))$ is a proof of $\rho^{-1}(\rho(\phi))$ in {\bf 2WKF}. As $\phi\equiv\rho^{-1}(\rho(\phi))$ is provable in {\bf 2WKF}, $\phi$ is also provable.
    \item By induction on the complexity of formulas, the only nontrivial case is the modal formulas. For example, let $\phi=\wnbm{c}\psi$. Then, $\mathfrak{M}, g\models_{s_1}\phi$ iff $\vartriangle_g\!\!(|\psi|)\geq c$ iff $\inf_{m\in|\psi|}\tilde{I}(g,m)\geq c$ iff $\inf_{m\not\in|\neg\rho_2(\psi|)}(1-\tilde{J}(g,m))\geq c$ (by the inductive hypothesis and the definition of $\tilde{J}$) iff  $\overline{\mathfrak{M}}, g\models_{s_1}[c]_{\mathbf p}\neg\rho_2(\psi)$, i.e., $\overline{\mathfrak{M}}, g\models_{s_1}\rho_1(\phi)$. 
    \item It follows from 2 immediately because the mapping of fuzzy context models to their complements is bijective. \epf
\end{enumerate}

The completeness of {\bf 2WKF} then follows from that of {\bf 2WKB} by the correspondence theorem (1) and (3).

\section{Extension to Multi-Relational Fuzzy Contexts}\label{sec6}
In this section, we demonstrate the possibility of extending the logic to reasoning with Boolean combinations of relations in multi-relational fuzzy contexts. 
  
In classical modal logic, when both necessity and sufficiency modalities are available in a logical language, it can be extended to express all Boolean combinations of multiple modalities. This leads to Boolean modal logic in ~\cite{Gargov1987}. Semantically, each primitive modality in Boolean modal logic is interpreted with respect to a binary accessibility relation in Kripke models and Boolean combinations of modalities correspond to the set-theoretic Boolean algebra of multiple relations. In particular, sufficiency modalities correspond to complements of binary relations.

Analogously, we can also extend {\bf 2WML} to a kind of two-sorted weighted Boolean modal logic ({\bf 2WBML}). Let $\Pi_0$ be a set of primitive modality indices and let $cl(\Pi_0)$ be its inductive closure over union, intersection, and complement. Formally, an index in $cl(\Pi_0)$ is defined by
\[\mathsf{i}::=\mathsf{0}\mid\mathsf{a}\mid\mathsf{i}\cap \mathsf{i}\mid\mathsf{i}\cup \mathsf{i}\mid\overline{\mathsf{i}},\] where $\mathsf{a}\in\Pi_0$.
Then, we modify the signature of {\bf 2WML} by attaching to each modality an index from $cl(\Pi_0)$. Thus, for example, we have a {\bf 2WBML} formula of the form $[c]_{\bf o}^{\mathsf{i}\cap \mathsf{j}}\varphi$. Semantically, these formulas are interpreted in multi-relational fuzzy context models $\mathfrak{M}=(G,M,(\tilde{I}_{\mathsf i})_{{\mathsf i}\in cl(\Pi_0)},v)$, where $G,M$, and $v$ are defined as in fuzzy context models, and each $\tilde{I}_{\mathsf i}:G\times M\ra[0,1]$ is a fuzzy binary relation such that
for any $g\in G$ and $m\in M$
\begin{enumerate}
    \item $\tilde{I}_{\mathsf 0}(g,m)=0$
    \item $\tilde{I}_{\mathsf{i}\cap\mathsf{j}}(g,m)=\min(\tilde{I}_{\mathsf{i}}(g,m),\tilde{I}_{\mathsf{j}}(g,m))$
    \item $\tilde{I}_{\mathsf{i}\cup\mathsf{j}}(g,m)=\max(\tilde{I}_{\mathsf{i}}(g,m),\tilde{I}_{\mathsf{j}}(g,m))$
    \item $\tilde{I}_{\overline{\mathsf{i}}}(g,m)=1-\tilde{I}_{\mathsf{i}}(g,m)$.
\end{enumerate}
To obtain axiomatization of {\bf 2WBML}, we simply  replace modalities in the {\bf 2WML} system with indexed ones and add the extra axioms and rule as follows:
\begin{enumerate}
    \item Axioms
    \begin{enumerate}
        \item (DefU): $[x]^{{\mathsf i}\cup\mathsf{j}}\phi\equiv([x]^{\mathsf i}\phi\wedge[x]^{\mathsf j}\phi)$
        \item (DefI): $\wnb{x}^{{\mathsf i}\cap\mathsf{j}}\phi\equiv(\wnb{x}^{\mathsf i}\phi\wedge\wnb{x}^{\mathsf j}\phi)$
        \item (DefC): $ [x]^{\overline{\mathsf{i}}}\phi\equiv\wnb{x}^{\mathsf i}\neg\phi$
        \item (Def{\sf 0}): $[1]^{\mathsf{0}}\phi$
    \end{enumerate}
    \item Rule (EQ): $\vdash [x]^{\mathsf i}\phi\equiv[x]^{\mathsf j}\phi$ if $\vdash_{ZA}\mathsf{i}=\mathsf{j}$
\end{enumerate}
for $x\in\{c,c^+\}$, where $ZA$ is the set of axioms for Zadeh algebra. That is, if $BA$ is the set of Boolean algebra axioms, then $ZA=BA-\{\mathsf{i}\cup\overline{\mathsf{i}}=\overline{\mathsf{0}}, \mathsf{i}\cap\overline{\mathsf{i}}=\mathsf{0}\}\cup\{\overline{\overline{\mathsf{i}}}=\mathsf{i}\}$. We denote these axioms and rule as $\bf BM$.  

Note that axioms (DefU) and (DefI) are mutually derivable given axiom (DefC) and the rule (EQ). In fact, (DefC) and(EQ) are powerful enough to eliminate one of the necessity or sufficiency fragments in the {\bf 2WML} system. The axioms for necessity and  sufficiency modalities in {\bf 2WML} also become mutually derivable via the connection of (DefC) and (EQ). Hence, we can remove one of them if the independence of axioms is an objective of the axiomatization. More specifically, let us still use the same naming conventions to denote the indexed versions of the axiomatic system {\bf 2WML} and  its fragments, axioms and rules. Then, we can define
\begin{enumerate}
    \item {\bf 2WBML}={\bf 2WML}+{\bf BM},
    \item {\bf 2WBML}$^N$={\bf 2WKB}+{\bf BM}+$\{$(CON$_1$),(CON$_2$)$\}$-$\{$(DefI)$\}$, and
    \item {\bf 2WBML}$^\vartriangle$={\bf 2WKF}+{\bf BM}+$\{$(CON$_1$),(CON$_2$)$\}$-$\{$(DefU)$\}$,
\end{enumerate}
and prove the following result.
\begin{proposition}{\rm {\bf 2WBML}, {\bf 2WBML}$^N$, and  {\bf 2WBML}$^\vartriangle$ are equivalent.}    
\end{proposition}

Finally, we note that (EQ) is an unorthodox rule since its premises relies on an external reasoning mechanism, i.e., the equational reasoning on a set of axioms for an algebra.

\section{Concluding Remarks}\label{sec7}
We have introduced a 2-sorted weighted modal logic for reasoning with fuzzy formal contexts. Our illustrative examples demonstrate its potential applicability to the representation of knowledge discovered from fuzzy databases. As the logic is  at the early stage of development yet, there are a lot of open issues to be addressed. First, the most pressing issue is the  completeness of the axiomatization for {\bf 2WML} and {\bf 2WBML}. Second, while {\bf 2WML} is still a two-valued logic, more generally, we may also study genuine many-valued modal logic following the mathematical fuzzy logic paradigm initiated by H\'{a}jek~\cite{haj}. In addition, considering the frequency of connections between objects and attributes, we can also study probabilistic reasoning~\cite{ogn} about formal contexts. Last, to apply the formalism to practical data mining problem, we will need to implement its automated reasoning system and test it on some real datasets.

\subsubsection{\ackname} 
This work is partially supported by the National Science and Technology Council of Taiwan under Grants: NSTC 113-2221-E-001-018-MY3 and NSTC 113-2221-E-001-021-MY3. 
\bibliographystyle{splncs04}
\bibliography{name}
\newpage
\appendix
\section{Properties of Fuzzy Contexts and Cut-Based Concepts}
In this appendix, we prove that pairs of operators defined in (\ref{fop}) and (\ref{rop}) form Galois connections and investigate some basic properties of cut-based concepts. We start with the definition of a Galois connection \cite{textfca}.
\begin{definition}
    {\rm Let $\phi:P\rightarrow Q$ and $\psi:Q\rightarrow P$ be maps between two ordered sets $(P, \leq)$ and $(Q, \leq )$. Then, such a pair of maps is called a Galois connection between the  ordered sets if  for all $p_{1}, p_{2}\in P$ and $q_{1}, q_{2}\in Q$,
     \begin{enumerate}
         \item  $p_{1}\leq p_{2}$ implies $\phi p_{1}\geq \phi p_{2}$.
         \item $q_{1}\leq q_{2}$ implies that $\psi q_{1}\geq \psi q_{2}$.
         \item $p\leq \psi\phi p$ and $q\leq \phi\psi q$.
     \end{enumerate}

    }
\end{definition}

\noindent The proposition below gives an alternative definition of  Galois connection.
\begin{proposition}
    \label{galaltdef}{\rm A pair $(\phi, \psi)$ of maps between  two ordered sets $(P, \leq)$ and $(Q, \leq)$ is a Galois connection if and only if $p\leq\psi q$ if and only if $q\leq \phi p$.}
\end{proposition}
Then, we can prove the following lemma.
\begin{mylma}
		\label{gloisp}
		{\rm Let $(G,M, \tilde{I}) $ be a fuzzy formal context and let $A, A_{1}, A_{2}\subseteq G$ and $B, B_{1}, B_{2}\subseteq M$. Then, the following holds for any $c\in[0,1]$.
			\begin{enumerate}
				\item If $A_{1}\subseteq A_{2}$ then $(A_{2})_{c}^{+}\subseteq (A_{1})_{c}^{+}$.  and $B_{1}\subseteq B_{2}$ implies that $(B_{2})^{-}_{c}\subseteq (B_{1})^{-}_{c}$.
				\item $A\subseteq (A^{+}_{c})^{-}_{c}$ and $B\subseteq (B^{-}_{c})^{+}_{c}$
				\item  $A^{+}_{c}= ((A^{+}_{c})^{-}_{c})^{+}_{c}$. and  $B^{-}_{c}= ((B^{-}_{c})^{+}_{c})^{-}_{c}$
		\end{enumerate}}
	\end{mylma}
    {\pf}
We prove the case fo subsets of $G$, and the other case can be proved in a similar way.\\
	1. As $A_{1}\subseteq A_{2}$, $A^+_{2}(m)\leq A^+_{1}(m)$ for all $m\in M$ by Proposition \ref{prop1}(1). Hence, for and $m\in M$ and $c\in[0,1]$, $A^+_{2}(m)\geq c$ implies $A^+_{1}(m)\geq c$, that is, $(A_{2})^{+}_{c}\subseteq (A_{1})^{+}_{c}$ by the definition of $c$-cut. \\
	2. For $g\in G$, $g\in ((A)^{+}_{c})^{-}_{c}$ if and only if $\inf_{m\in (A)_{c}^{+}}\tilde{I}(g,m)\geq c$.  Now $m\in (A)^{+}_{c}$ if and only if $\inf_{g\in A}\tilde{I}(g, m)\geq c$. Thus, if $g_{0}\in A$ then for all $m\in (A)^{+}_{c} $, $\tilde{I}(g_{0},m)\geq c$ which implies that $\inf_{m\in (A)_{c}^{+}}\tilde{I}(g_{0},m)\geq c$. Hence $g_{0}\in ((A)^{+}_{c})^{-}_{c}$ whence $A\subseteq ((A)^{+}_{c})^{-}_{c} $.\\
	3. Applying 1 to the result in 2, it follows that  $ ((A^{+}_{c})^{-}_{c})^{+}_{c}\subseteq A^{+}_{c}$. Replacing $B$ with $A^{+}_{c}$ in 2, we have $A^{+}_{c}\subseteq ((A^{+}_{c})^{-}_{c})^{+}_{c}$. Hence $A^{+}_{c}= ((A^{+}_{c})^{-}_{c})^{+}_{c}$.
	{\epf}
    
With this lemma, we can prove the following theorem.
	\begin{theorem}
		\label{gloisfuzzy}{\rm For a fuzzy formal context $(G, M, \tilde{I})$, the pair of operators defined in (\ref{fop}) forms a Galois connection.}
	\end{theorem}
Next, we have the following two propositions.
	\begin{proposition}
		\label{relabetwoperators}
		{\rm For  a fuzzy formal context $(G, M, \tilde{I})$, we define its complement as $(G, M, \tilde{J})$ where $\tilde{J}:G\times M\rightarrow [0,1]$, $\tilde{J}(g, m)=1-\tilde{I}(g, m)$ for all $(g,m)\in G\times M$. Then, for $A\subseteq G$ and $B\subseteq M$ the following holds.
\begin{enumerate}
\item $A^{+}_{\tilde{I}}=A'^{\Box}_{\tilde{J}}$ and ${B^{-}_{\tilde{I}}}={B'^{\Box}_{\tilde{J}}}$

\item $A^\Box_{\tilde{I}}=A'^+_{\tilde{J}}$ and $B^\Box_{\tilde{I}}=B'^-_{\tilde{J}}$

\item $(A^+_{\tilde{I}})_c=(A'^\Box_{\tilde{J}})_c$ and $(B^-_{\tilde{I}})_c=(B'^\Box_{\tilde{J}})_c$

\item $(A^\Box_{\tilde{I}})_c=(A'^+_{\tilde{J}})_c$ and $(B^\Box_{\tilde{I}})_c=(B'^-_{\tilde{J}})_c$

\item $A^\Diamond_{>c}=(A'^\Box_{1-c})'$ and  $B^\Diamond_{>c}=(B'^\Box_{1-c})'$.
				
\item  $(A^\Diamond_{\tilde{I
				 }})_{>c}=((A^+_{\tilde{J}})_{1-c})'$ and  $(B^\Diamond_{\tilde{I}})_{>c}=((B^-_{\tilde{J}})_{1-c})'$.
		\end{enumerate}
Here, we use $S'$ to denote the complement of a subset $S$. }

	\end{proposition}
	{\pf}
We only prove the case for $A$ in each item and the other case is similar. \\
\noindent 1. By Proposition \ref{prop1}, for any $m\in M$, $A^{+}_{\tilde{I}}(m)=\inf_{g\in A}\pi_m(g)=\inf_{g\notin A'}\tilde{I}(g,m)=\inf_{g\notin A'}1-\tilde{J}(g,m)=A'^\Box_{\tilde{J}}(m)$.\\  
\noindent 2. It follows immediately from 1 by exchanging $\tilde{I}$ with $\tilde{J}$ and $A$ with $A'$.\\
\noindent 3. and 4.  follows from 1 and 2, respectively, by the definition of $c$-cut.\\
\noindent 5. For any $m\in M$, we have
\begin{eqnarray*}
 m\in(A'^{\Box}_{1-c})'& {\rm iff} & m\notin A'^{\Box}_{1-c}\\
 & {\rm iff} & A'^{\Box}(m)\not{\geq} 1-c\\
  & {\rm iff} & \inf_{g\notin A'} 1-\tilde{I}(g,m)\not{\geq} 1-c\\
  & {\rm iff} & \exists g_0\in A, 1-c> 1-\tilde{I}(g_{0},m)\\
  &{\rm iff}& A^\Diamond(m)=\sup_{g\in A}\tilde{I}(g, m)\geq \tilde{I}(g_{0}, m)> c\\
  &{\rm iff}& m\in (A^\Diamond)_{>c}
\end{eqnarray*}
\noindent 6. It follows from 4 and 5.
    {\epf}
\begin{proposition}
		\label{modalglois}
		{\rm Let $(G,M, \tilde{I})$ be a fuzzy formal context and $A, A_{1}, A_{2}\subseteq G$ and $B, B_{1}, B_{2}\subseteq M $. Then the following holds.
			\begin{enumerate}
				\item If $A_{1}\subseteq A_{2}$ then $(A_1^\Box)_{c}\subseteq (A_2^\Box)_{c}$ and $(A_1^\Diamond)_{>c}\subseteq (A_2^\Diamond)_{>c} $.
				\item If $B_{1}\subseteq B_{2}$ then $(B_1^\Box)_{c}\subseteq (B_{2}^\Box)_{c} $ and $(B_{1}^\Diamond)_{>c}\subseteq (B_{2}^\Diamond)_{>c}$.
				\item  $A\subseteq (A^\Diamond_{>1-c})^\Box_{c}$ and $B\subseteq (B^\Diamond_{>1-c})^\Box_{c}$
				\item 
				$(A^\Box_{c})^\Diamond_{>1-c}\subseteq A$ and $(B^\Box_{c})^\Diamond_{>1-c}\subseteq B$
		\end{enumerate}}
	\end{proposition}
	{\pf}
    \noindent 1. By Proposition~\ref{prop1}, $A_{1}\subseteq A_{2}$ implies that $A_1^\Box(m)\leq A^\Box_{2}(m)$ and $A_1^\Diamond(m)\leq A^\Diamond_{2}(m)$ for any $m\in M$ as both necessity and possibility are monotonic measures. Then, the results follows by the definition of $c$-cuts and strict $c$-cuts.\\
	\noindent 2. The proof is similar to that for 1.\\
	\noindent 3. By Lemma \ref{gloisp}(2), $A\subseteq (((A^+_{\tilde{J}})_c)^-_{\tilde{J}})_c$. By Proposition \ref{relabetwoperators}(6), $A\subseteq ((((A^\Diamond_{\tilde{I}})_{>1-c})^{\prime})^-_{\tilde{J}})_{c}$ and again by Proposition \ref{relabetwoperators}(4), $A\subseteq (((A^\Diamond_{\tilde{I}})_{>1-c})^\Box_{\tilde{I}})_{c}$. Omitting the subscript $\tilde{I}$, we have $A\subseteq (A^\Diamond_{>1-c})^\Box_{c}$.  
    
    \noindent 4. By Lemma \ref{gloisp}(2), $A'\subseteq (((A'^+_{\tilde{J}})_{c})^-_{\tilde{J}})_{c}$. By Proposition \ref{relabetwoperators}(4), $A'\subseteq (((A^\Box_{\tilde{I}})_c)^-_{\tilde{J}})_{c}$ and again by Proposition \ref{relabetwoperators}(6) $A'\subseteq ((((A^\Box_{\tilde{I}})_c)^\Diamond_{\tilde{I}})_{>1-c})'$ which implies that\\ $ ((((A^\Box_{\tilde{I}})_c)^\Diamond_{\tilde{I}})_{>1-c})\subseteq A$.  Omitting the subscript  $\tilde{I}$, we have $(A^\Box_{c})^\Diamond_{>1-c}\subseteq A$.
	{\epf}

Based on these two propositions, we can prove the following theorem.	
	\begin{theorem}
		{\rm For a fuzzy formal context $(G, M, \tilde{I})$, the pair of operators defined in (\ref{rop})  forms a Galois connection.}
	\end{theorem}
	{\pf}
	Let $A\subseteq G$ and $B\subseteq M$.  Suppose  $A^\Diamond_{>1-c}\subseteq B$. Then, by Proposition \ref{modalglois}(3) and (1), $A\subseteq (A^\Diamond_{>1-c})^\Box_{c}\subseteq B^\Box_{c}$. On the other hand,  if $A\subseteq B^\Box_{c}$, then by Proposition \ref{modalglois}(1) and (4) $A^\Diamond_{>1-c}\subseteq (B^\Box_c)^\Diamond_{>1-c} \subseteq B$. Hence, the result follows  from  Proposition \ref{galaltdef}. 
	{\epf} 

Regarding the properties of cut-based concepts, we first prove the following theorem.
	
	\begin{theorem}
		\label{relationbconcept}
		{\rm Let $\tilde{\mathbb{K}}=(G,M, \tilde{I})$ be a fuzzy formal context and let $\overline{\tilde{\mathbb{K}}}=(G, M, \tilde{J})$ be its complemented fuzzy context as defined in Proposition~\ref{relabetwoperators}. Then, the following hold.
			\begin{enumerate}
			\item $(A, B)\in \mathbf{B}_{c}(\tilde{\mathbb{K}})$ iff $(A', B)\in \mathbf{O}_{c}(\overline{\tilde{\mathbb{K}}})$.
				\item $(A, B)\in \mathbf{B}_{c}(\tilde{\mathbb{K}})$ iff $(A, B')\in \mathbf{P}_{c}(\overline{\tilde{\mathbb{K}}})$.
				\item $(A, B)\in\mathbf{O}_{c}(\tilde{\mathbb{K}})$ if and only if $(A', B')\in \mathbf{P}_{c}(\overline{\tilde{\mathbb{K}}})$.
		\end{enumerate}}
	\end{theorem}
	{\pf}
	The proofs of (1) and (2) are similar , and (3) follows from them immediately. We simply present the proof of (1) as an example. Let $(A, B)\in \mathbf{B}_{c}(\tilde{\mathbb{K}})$. Then  $(A^+_{\tilde{I}})_{c}=B$ and $(B^-_{\tilde{I}})_{c}=A$. By Proposition \ref{relabetwoperators}(3) and (6),  $(A
    ^{\prime\square}_{\tilde{J}})_{c}=B$ and $(A^{\lozenge}_{\tilde{J}})_{>1-c}=A'$. Hence $(A', B)\in \mathbf{O}_{c}(\overline{\tilde{\mathbb{K}}})$.
	{\epf}

The following results are analogous to those in the classical case. We simply state them without the proofs as they closely follow the classical ones.
\begin{theorem}
		\label{fuzyconcept}
		{\rm For a fuzzy formal context $(G, M, \tilde{I})$, define join and meet in $\mathbf{B}_{c}(\tilde{\mathbb{K}})$, $\mathbf{O}_{c}(\tilde{\mathbb{K}})$, and $\mathbf{P}_{c}(\tilde{\mathbb{K}})$ respectively by 
\[\begin{array}{ll}
 (A, B)\vee (C, D):=((B\cap D)^{-}_{c}, B\cap D)    & (A, B)\wedge (C, D):=(A\cap C, (A\cap C)^{+}_{c}), \\
 (A, B) \vee (C, D) := ((B \cap D)^{\lozenge}_{>1-c}, B\cap D)    & (A, B) \wedge (C, D) := (A \cap C, (A \cap C)^{\square}_{c}),\\
(A, B) \vee (C, D) := ((B \cap D)^{\square}_{c}, B\cap D)&(A, B) \wedge (C, D) := (A \cap C, (A \cap C)^{\lozenge}_{>1-c}).
\end{array}
\] 
Then, they all form complete lattices.}
	\end{theorem}

\begin{proposition}{\rm 
 \begin{enumerate}
     \item The $c$-formal concept lattice $\underline{\mathbf{B}}_{c}(\tilde{\mathbb{K}})$ is dually isomorphic to the $c$-object oriented concept lattice $\underline{\mathbf{O}}_{c}(\overline{\tilde{\mathbb{K}}})$.
     \item The $c$-formal concept lattice $\underline{\mathbf{B}}_{c}(\tilde{\mathbb{K}})$ is isomorphic to the $c$-property oriented concept lattice $\underline{\mathbf{P}}_{c}(\overline{\tilde{\mathbb{K}}})$.
     \item the $c$-object oriented concept lattice $\underline{\mathbf{O}}_{c}(\tilde{\mathbb{K}})$ is dually isomorphic to the $c$-property oriented concept lattice $\underline{\mathbf{P}}_{c}(\overline{\tilde{\mathbb{K}}})$.
 \end{enumerate}}  
\end{proposition}

\section{Proof of Theorem~\ref{soucomkbwml}} 
\subsection{Proof of Lemma~\ref{bcondition}}
Let  $\Sigma_g$  and $\Sigma_m$ be maximally consistent subsets of ${\cal L}_{s_1}(D)$ and ${\cal L}_{s_2}(D)$, respectively. Then, we have
\begin{enumerate}
    \item $\Sigma_g/[c]_{\mathbf p}\subseteq\Sigma_m$ iff $\Sigma_m/[c]_{\mathbf o}\subseteq\Sigma_g$,
    \item $\Sigma_g/[c^+]_{\mathbf p}\subseteq\Sigma_m$ iff $\Sigma_m/[c^+]_{\mathbf o}\subseteq\Sigma_g$.
\end{enumerate}
{\pf}\begin{enumerate}
    \item Assume that $\Sigma_g/[c]_{\mathbf p}\subseteq\Sigma_m$ but $\Sigma_m/[c]_{\mathbf o}\not\subseteq\Sigma_g$. Then, there exists $\varphi\in\Sigma_m/[c]_{\mathbf o}$ and $\varphi\not\in\Sigma_g$. By (Bookkeeping) axiom, $[c]_{\mathbf o}\varphi\in\Sigma_m$, and by maximal consistency, $\neg\varphi\in\Sigma_g$. Then, by $(B_{[c]})$ and (MP), $[c]_{\mathbf p}\neg[c]_{\mathbf o}\neg\neg\varphi\in\Sigma_g$, and thus $[c]_{\mathbf p}\neg[c]_{\mathbf o}\varphi\in\Sigma_g$ by further propositional modal reasoning, which implies $\neg[c]_{\mathbf o}\varphi\in\Sigma_g/[c]_{\mathbf p}\subseteq\Sigma_m$, in contradiction with the consistency of $\Sigma_m$. The other direction is proved similarly using the second item of the $(B_{[c]})$ axiom.
    \item Analogously, we can prove this using the $(B_{[c^+]})$ axiom.\epf
\end{enumerate}

\subsection{Proof of Model Existence Lemma}
There exists a fuzzy binary relation $\tilde{I}:G\times M\rightarrow[0,1]$ satisfying the two conditions for the definition of canonical models. \\  
{\pf} The result relies on that $D$ is finite. Let us assume that the elements of $D$ are enumerated decreasingly as $1=c_1>c_2>\cdots>c_k=0$. Then, for any $\Sigma_g\in G$, we have $\Sigma_g/[c_1^+]_{\mathbf p}=\emptyset$ and for any  $1\leq i<k$,
\[\Sigma_g/[c_i^+]_{\mathbf p}\subseteq \Sigma_g/[c_i]_{\mathbf p}\subseteq \Sigma_g/[c_{i+1}^+]_{\mathbf p}.\] Hence, for any $\Sigma_g\in G$ and $\Sigma_m\in M$ , $\tilde{I}(\Sigma_g,\Sigma_m)$ can be defined in the following way:
\begin{enumerate}
\item if $\Sigma_g/[c_i^+]_{\mathbf p}\subseteq\Sigma_m\not\supseteq \Sigma_g/[c_i]_{\mathbf p}$ for some $1\leq i<k$, then $\tilde{I}(\Sigma_g,\Sigma_m)=1-c_i$
\item if $\Sigma_g/[c_i]_{\mathbf p}\subseteq\Sigma_m\not\supseteq \Sigma_g/[c_{i+1}^+]_{\mathbf p}$ for some $1\leq i<k$, then $\tilde{I}(\Sigma_g,\Sigma_m)=1-c$ for some $c\in(c_i,c_{i+1})$
\item if $\Sigma_g/[c_k^+]_{\mathbf p}\subseteq\Sigma_m$, then $\tilde{I}(\Sigma_g,\Sigma_m)=1$.
\end{enumerate}
Then, it is easily verified that $\tilde{I}$ satisfies the requirements.
\epf
\subsection{Proof of Truth Lemma}
Let  ${\mathfrak M}=(G,M,\tilde{I},v)$ be a ${\mathcal L}(D)$-canonical model and let $\varphi\in{\mathcal L}_{s_1}(D)$ and $\psi\in{\mathcal L}_{s_2}(D)$. Then, for any $\Sigma_g\in G$ and $\Sigma_m\in M$, we have
\begin{enumerate}
    \item ${\mathfrak M},\Sigma_g\models_{s_1}\varphi$ iff $\varphi\in\Sigma_g$ and
    \item ${\mathfrak M},\Sigma_m\models_{s_2}\psi$ iff $\psi\in\Sigma_m$. 
\end{enumerate}
{\pf} We prove the lemma by simultaneous induction on the complexity of both sorts of formulas. \\
\noindent Induction base: this follows immediately from the definition of $v$.\\
\noindent Induction step: the proofs for Boolean connectives  are straightforward. For the modal formulas, we only show the case of $s_1$-sorted ones. The other one is similar.  Now, let us consider the case of $\varphi=[c]_{\mathbf p}\psi$. If $c=0$, then the result trivially holds by the (Bookkeeping) axiom (v). Hence, we can assume that $c>0$.  

For one direction, assume that $\varphi\in\Sigma_g$. Then $\psi\in \Sigma_g/[c]_{\mathbf p}$. Thus, $\psi\in\Sigma_m$ for any $\Sigma_m$ such that $\tilde{I}(\Sigma_g,\Sigma_m)>1-c$ by the first condition of the fuzzy relation in a canonical model. By the inductive hypothesis, for any $\Sigma_m$ such that $\tilde{I}(\Sigma_g,\Sigma_m)>1-c$, ${\mathfrak M},\Sigma_m\models_{s_2}\psi$, which implies that $\Pi_g(|\neg\psi|)\leq 1-c$, i.e., $N_g(|\psi|)\geq c$. Therefore, we have ${\mathfrak M},\Sigma_g\models_{s_1}\varphi$. 

For the other direction, assume that $\varphi\not\in\Sigma_g$. Then, $\Sigma_g/[c]_{\mathbf p}\cup\{\neg\psi\}$ is consistent. Otherwise, there exist $\psi_1,\psi_2,\cdots,\psi_n\in \Sigma_g/[c]_{\mathbf p}$ such that $\vdash_{s_2}\psi_1\rightarrow(\psi_2\rightarrow(\cdots(\psi_n\rightarrow\psi)))$ and then by (UG$_{[1]}$), (Bookkeeping), (K$_[c]$), and (MP), we have $\Sigma_g\vdash_{s_1}\varphi$, which implies $\varphi\in\Sigma_g$ by the maximal consistency of $\Sigma_g$ and is contradictory with the assumption. Hence, we can find a maximally consistent $\Sigma_m\in M$  such that $\Sigma_g/[c]_{\mathbf p}\cup\{\neg\psi\}\subseteq\Sigma_m$. Again, by the condition on the canonical model and inductive hypothesis, this means that there exists $\Sigma_m\in M$ such that $\tilde{I}(\Sigma_g,\Sigma_m)>1-c$ and ${\mathfrak M},\Sigma_m\not\models_{s_2}\psi$. Thus, we have $\Pi_g(|\neg\psi|)>1-c$ and $N_g(|\psi|)<c$, i.e.  ${\mathfrak M},\Sigma_g\not\models_{s_1}\varphi$.

The proof for the case of $\varphi=[c^+]_{\mathbf p}\psi$ is similar except that we have to use the second condition on the fuzzy relation of the canonical model. \epf

\end{document}